\documentclass[10pt,onecolumn,journal,letterpaper]{IEEEtran}
\usepackage[square,comma,sort&compress,numbers]{natbib}
\usepackage{algpseudocode}
\usepackage{amsmath}
\usepackage{amsfonts}
\usepackage{amssymb}
\usepackage{amsthm}
\usepackage{mathtools}
\usepackage[detect-weight=true,detect-mode=true]{siunitx}
\usepackage{nicefrac}
\usepackage{dsfont}
\usepackage{upgreek}
\usepackage{enumitem}
\usepackage{bbm}
\usepackage{booktabs}
\usepackage{graphicx}
\usepackage{subcaption}
\usepackage[hidelinks]{hyperref}

\begin{document}

\author{
	Annan~Dong*,
	Osvaldo~Simeone*$\dagger$,
	Alexander~M.~Haimovich*,
	and Jason~A.~Dabin** \\
	* CWiP, New Jersey Institute of Technology, USA, ad372,haimovich@njit.edu\\
	$\dagger$ King's College London, UK, osvaldo.simeone@kcl.ac.uk\\
	** Naval Information Warfare Center Pacific, San Diego (CA), USA 
    \thanks{The work of A.\ M.\ Haimovich is supported in part by the Booz Allen Hamilton Inc. under agreement No.\ 12-D-7248 TO 0046. 
    	
    The work of O. Simeone was supported by the European Research Council (ERC) under the European
    Union’s Horizon 2020 research and innovation programme (grant agreement No. 725731).}
}

\title{Blind Sparse Estimation of Intermittent Sources over Unknown Fading Channels}

\maketitle


\begin{abstract}
	
Radio frequency sources are observed at a fusion center via sensor measurements made over slow flat-fading channels. The number of sources may be larger than the number of sensors, but their activity is sparse and intermittent with bursty transmission patterns. To account for this, sources are modeled as hidden Markov models with known or unknown parameters. The problem of blind source estimation in the absence of channel state information is tackled via a novel algorithm, consisting of a dictionary learning (DL) stage and a per-source stochastic filtering (PSF) stage. The two stages work in tandem, with the latter operating on the output produced by the former.
Both stages are designed so as to account for the sparsity and memory of the sources. To this end, smooth LASSO is integrated with DL, while the forward-backward algorithm and Expectation Maximization (EM) algorithm are leveraged for PSF. It is shown that the proposed algorithm can enhance the detection and the estimation performance of the sources, and that it is robust to the sparsity level.
	
\end{abstract}

\begin{IEEEkeywords}
	Blind source separation, Wireless networks, Dictionary learning, Intermittent and sparse sources, Hidden Markov model
\end{IEEEkeywords}

\section{Introduction}

Blind source separation (BSS) refers to the separation of a set of source signals from a set of mixed signals, without resorting to any a priori information about the source signals or the mixing process \cite{BSShandbook}. BSS exploits only the information carried by the received signals themselves, hence the term \textit{blind}. BSS has numerous applications in speech recognition \cite{Speech03,BD11}, image extraction \cite{Image02,Image97}, and surveillance \cite{Surv00,Surv04}. Different metrics are used to evaluate the performance of BSS methods depending on the applications. For example, signal-to-interference ratio is used in \cite{Speech03} for speech separation, and a performance index is introduced in \cite{Image02} for image feature extraction. Based on these metrics, many approaches have been proposed to solve BSS problems, such as independent component analysis (ICA) \cite{ICA00}, principal component analysis (PCA) \cite{PCA87}, and singular value decomposition (SVD) \cite{SVD70}.

This paper addresses BSS in wireless networks. We are specifically interested in the set-up illustrated in Fig.~\ref{fig:SM}, in which a fusion center observes a number of radio sources via noisy sensor measurements over unknown channels. The system may model an Internet-of-Things (IoT) system, such as LoRa, Sigfox, or Narrow Band-IoT (NB-IoT) \cite{IoT1,IoT2}.

In wireless networks involving multiple terminals operating over flat fading channels, the signals received at a terminal are linear mixtures of the signals emitted by the transmitting terminals. The need for BSS arises in non-collaborative applications in which the signals and the channels through which they are received at a terminal are both unknown.
ICA has been widely applied to solve BSS problems in wireless networks \cite{Alavi16,ICACRN11,ICAMIMO04,ICAOFDM07}, since it yields a useful decomposition with only scaling, and permutation ambiguities \cite{ICA}. To achieve signal separation, ICA relies on the statistical independence and on the non-Gaussian distribution of the components of the mixture. Key assumptions made in the implementation of the various forms of ICA are that the underlying mixing process has the same number of inputs and outputs, and that all sources are active throughout the observation interval. These assumptions are limiting and not suitable for the applications under study in this work, as discussed next.

With the aim of capturing key aspects of IoT systems, this paper focuses on practical wireless scenarios in which the number of latent sources is generally larger than the number of sensors, but the sources are active intermittently with bursty transmission patterns. 
The cumulative time a source is active is a small fraction of the overall observation time, and the sources' on-off patterns vary slowly. In order to capture these properties, the sources are modeled as hidden Markov processes with known or unknown parameters. Source memory will be seen to be instrumental in enabling source separation.
The wireless network operates over slow flat-fading channels; sensors communicate with a fusion center over ideal channels; and all nodes are time-synchronized to the same clock by the fusion center.

The general BSS problem with more sources than sensors can be formulated as an underdetermined linear system $\mathbf{A}\mathbf{x}=\mathbf{y}$ in the absence of noise. The columns of the matrix $\mathbf{A}$ serve as a basis for expressing the observations $\mathbf{y}$. The set of basis signals that form the matrix $\mathbf{A}$ is called a dictionary.
Underdetermined linear systems of equations of the form $\mathbf{A}\mathbf{x}=\mathbf{y}$ have infinitely many solutions when the matrix $\mathbf{A}$ is full rank. Regularization may introduce additional conditions on the solutions, for example favoring smaller values of $\mathbf{x}$, leading to unique solutions of the underdetermined linear system. Sparse representations for which the solution $\mathbf{x}$ is unique have been the subject of intensive research resulting in a large body of literature \cite{OMPbook,FOCUSS97,OOMP02}. Pre-defined dictionaries, such as based on Fourier transforms, are convenient and computationally fast, but in the BSS problem, both the dictionary $\mathbf{A}$ and the signals $\mathbf{x}$ are unknown.

Sparse representation problems for which the dictionary is unknown require \textit{dictionary learning} (DL) in addition to signal recovery \cite{DL10,DL11}.
The advantage of DL is to enable a system to learn a dictionary adaptively from a set of observations rather than assume a prescribed rigid dictionary.
Among other problems, DL methods have been applied to joint direction of arrival (DOA) estimation and array calibration \cite{DLWC17}, linear transceiver design \cite{DLWC13}, cloud K-singular value decomposition (K-SVD) for big, distributed data \cite{DLWC15}, and channel representation for frequency-division duplexing (FDD) massive MIMO system \cite{DLWC16}.
In these examples, the DL algorithms solve BSS problems in which the number of sources is larger than the number of sensors, but the methods are agnostic to time variability properties of sources with memory.
In \cite{HMM09} and \cite{HMM10}, the authors set up a hidden Markov model (HMM) to solve a BSS problem. However, in \cite{HMM09} the memory of the sources is not accounted for, while in \cite{HMM10} a simplified model is assumed whereby only one source can appear or disappear at any given time.

In this context, we propose a two-stage algorithm for solving the BSS problem for sources with memory modeled by an HMM and observed over slow flat-fading channels. The proposed algorithm comprises a DL stage and a Per-source Stochastic Filtering (PSF) stage. The DL stage of the proposed algorithm exploits knowledge about the source sparsity and memory to aid with the source separation. The effect of source memory is introduced by a penalty term that discourages solutions with short-duration transmissions by means of a smooth LASSO algorithm.
The input to the DL stage are observations from the sensors. The output from the DL stage are channel estimates, source signal estimates and source states (active or inactive). The source state estimates produced by the DL algorithm may be viewed as the output of a binary asymmetric channel in which some of the states are "flipped" with respect to the true states. The error probabilities associated with the flipped states are referred to as flipping probabilities. The PSF stage consists of a forward-backward step along with an Expectation Maximization (EM) step that estimate the unknown
HMM transition probabilities and flipping probabilities.

The main contributions of this paper are summarized as follows:
\begin{itemize}
	\item A two-stage architecture is introduced for solving the problem of blind source estimation of HMM sources over slow flat-fading channel. The sources feature intermittent activity, and the number of latent source may be larger than the number of sensors;
	\item A smooth DL algorithm is proposed that is capable of exploiting source memory to support channel estimation, signal estimation and source detection. Two simplified versions of the smooth DL algorithms are introduced to reduce the computational complexity;
	\item An PSF algorithm is introduced that is capable of operating in the absence of a priori information about the HMM parameters and the state estimation flipping probabilities. This algorithm integrates an forward-backward step with an EM step.
\end{itemize}

The rest of the paper is organized as follows. The system model and a hidden Markov source model are presented in Sec.~\ref{sec:SM}. Background on DL algorithms is provided in Sec.~\ref{sec:pre}. In Sec.~\ref{sec:SDL}, we propose a two-stage DL-based algorithm to solve the BSS problem for sources with memory in wireless networks. The PSF is described in detail in Sec.~\ref{sec:PSF} for the two cases with known and unknown source parameters. Finally in Sec.~\ref{sec:NR}, simulation numerical results are shown to support that our proposed algorithm can separate sources and recover source signals with higher accuracy than existing DL algorithms.

\section{System Model} \label{sec:SM}
	
\begin{figure}
	\centering
	\begin{minipage}[c]{0.7\columnwidth}
		\includegraphics[width=\columnwidth]{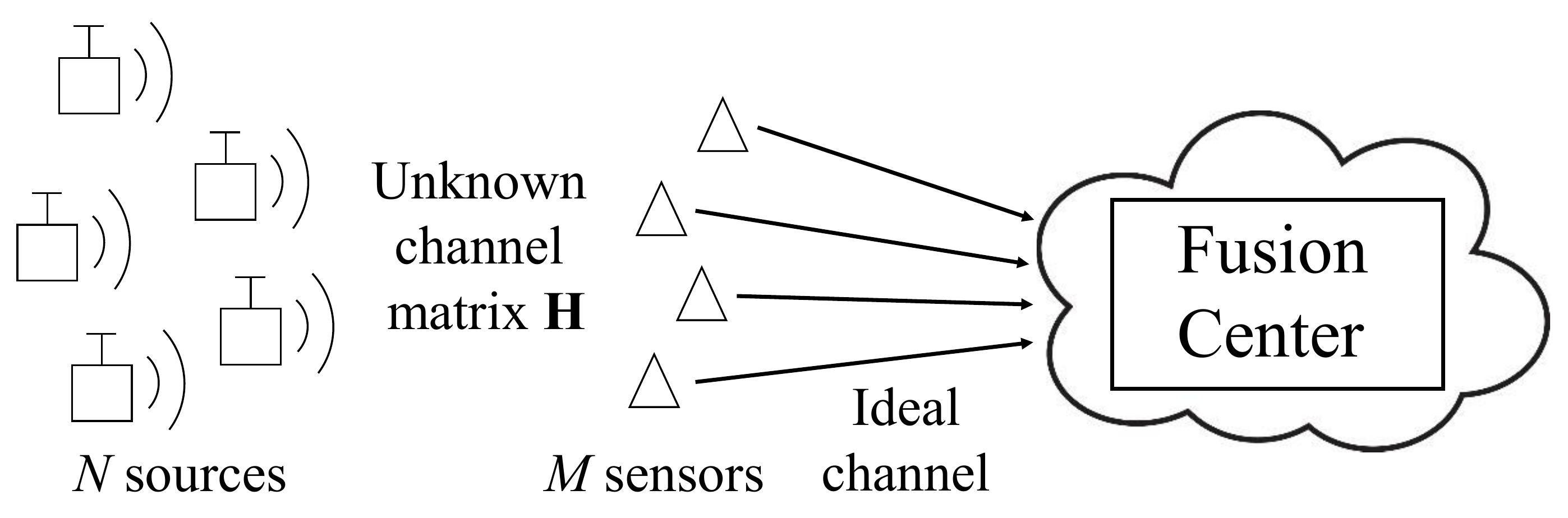}
		\caption{Wireless Sources transmit sporadically over a flat fading channel. The fusion center in the ``cloud" wishes to estimate the activity and transmitted signals based on the signals received by the distributed radio sensors.}
		\label{fig:SM}
	\end{minipage}
\end{figure}

Consider a system that includes $M$ receiving antennas, or radio sensors, and $N$ sources, as illustrated in Fig.~\ref{fig:SM}. The number of sources $N$ is generally larger than the number of receive antennas $M$. All sensors are connected to a fusion center, which may be implemented in a cloud processor, via backhaul links.
Equivalently, the fusion center has access to $N$ receive antennas.
Models with a fusion center reflect the architecture of IoT networks, such as LoRa, Sigfox, and NB-IoT \cite{IoT1,IoT2}. 
In our model, each source transmits intermittently, and hence is generally active only for a subset of the $T$ symbol periods over which the sensors collect data. 

Assuming that all nodes are time-synchronous, the discrete-time signal received by the $M$ sensors over $T$ symbol periods is given in matrix form as
\begin{equation} \label{eq:signal model}
	\mathbf{Y}=\mathbf{H}\mathbf{X} +\mathbf{Z},
\end{equation}
where $\mathbf{Y}=[\mathbf{y}(1),\dots,\mathbf{y}(T)]$ is an $M \times T$ matrix collecting as columns the $M \times 1$ received signals $\mathbf{y}(t)$ across all $T$ symbols $t=1,\dots,T$; $\mathbf{X}=[\mathbf{x}(1),\dots,\mathbf{x}(T)]$ is an $N \times T$ matrix that gathers the $N \times 1$ signals $\mathbf{x}(t)$ transmitted from all $N$ sources over time; $\mathbf{Z}=[\mathbf{z}(1),\dots,\mathbf{z}(T)]$ contains independent zero-mean complex Gaussian noise entries with variance $\sigma^2$; and $\mathbf{H}$ is the $M \times N$ complex fading channel matrix. The channel matrix $\mathbf{H}$ is assumed to be constant for $T$ symbol periods.

Given the intermittent nature of the traffic pattern of the sources, the $t$-th column vector $\mathbf{x}(t)$ that collects the $M$ symbols transmitted by the sources at time $t$, is generally sparse. In other words, only a subset of the entries of $\mathbf{x}(t)$ is non-zero. The signals $\mathbf{y}(t)$, for $t=1,\dots,T,$ are collected at the fusion center.
Based on the received signals $\mathbf{Y}$, the goal of the fusion center is to detect the sources' activity and to recover the signals $\mathbf{x}(t),$ for $t=1,\dots,T$, or equivalently the matrix $\mathbf{X}$, in the absence of information about the channel matrix $\mathbf{H}$.

For each source $n$, we define the activation pattern as a binary sequence $s_n(t)$, where the binary state $s_n(t)$ indicates whether a source is active or not. Specifically, when the state of source $n$ is $s_n(t)=1$, then the source is active, while it is inactive when $s_n(t)=0$. The binary state is assumed to be described by the following models.

\begin{figure}
	\centering
	\begin{minipage}[c]{0.7\columnwidth}
		\includegraphics[width=\columnwidth]{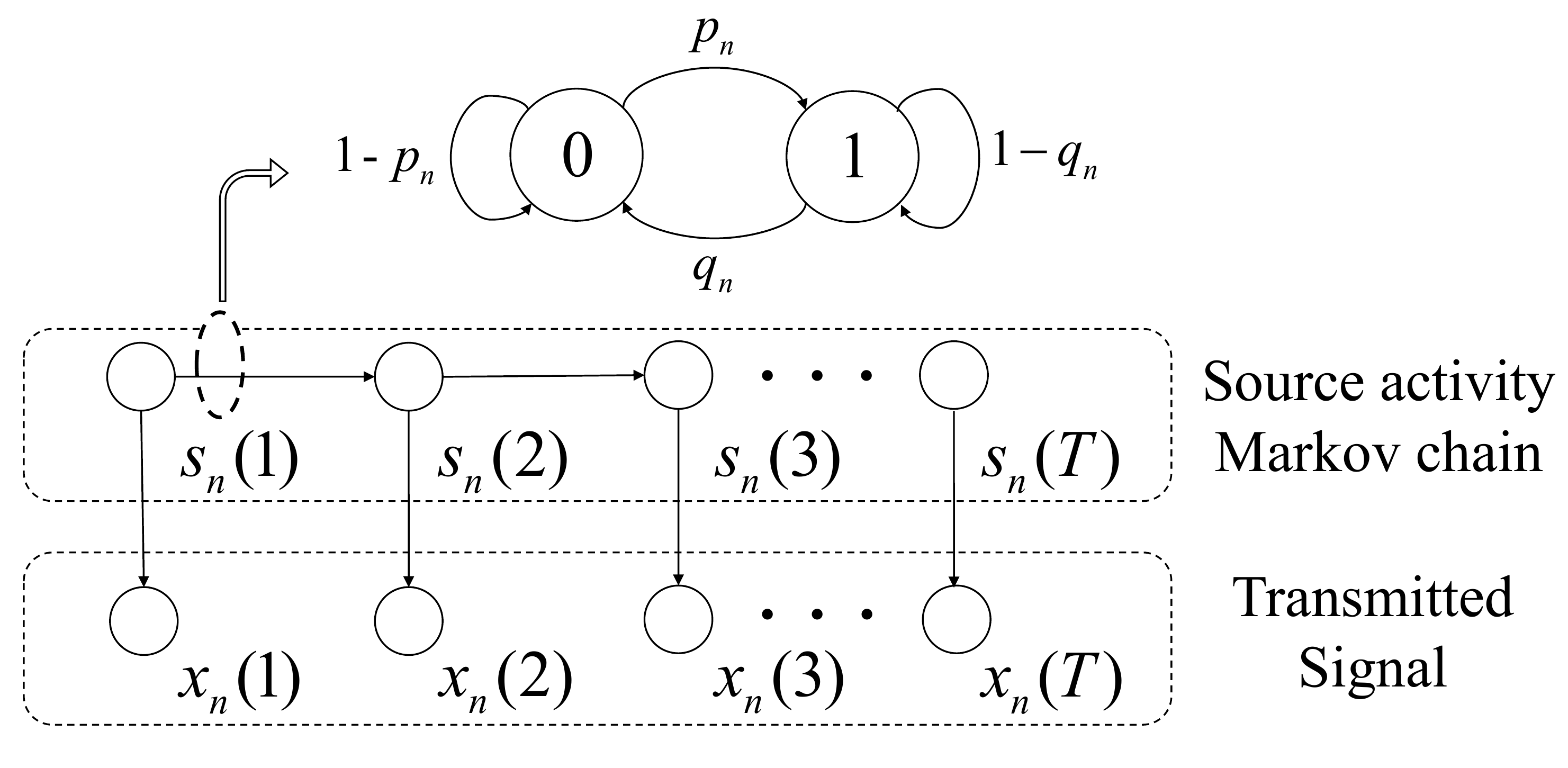}
		\caption{Hidden Markov Model (HMM) for a source $n$.}
		\label{fig:HMM}
	\end{minipage}
\end{figure}

\begin{enumerate}
	\item \textit{Intermittent and smooth deterministic model}: Each source is active for a small fraction of time, and its on-off patterns tend to have few switches between on and off states. As a result, the sequence $s_n(t)$ is ``smooth", i.e., it has a small number of transitions between on and off states;
	\item \textit{Probabilistic hidden Markov model}: As illustrated in Fig.~\ref{fig:HMM}, the activity $s_n(t)$ of each source $n$ follows a two-state Markov chain. The transition probabilities of the two-state Markov chain are defined as $p_n=\text{Pr}(s_n(t)=1|s_{n-1}(t)=0)$ and $q_n=\text{Pr}(s_n(t)=0|s_{n-1}(t)=1)$. We will consider both the cases in which the probabilities $p_n$ and $q_n$ are known and unknown.
\end{enumerate}

We emphasize that the two models are not mutually exclusive and can be assumed to hold simultaneously. In particular, we will propose to leverage the deterministic model in order to improve the performance of source separation and the stochastic model to refine the estimates obtained by sources separation.

When a source $n$ is active, i.e., when $s_n(t)=1$, it transmits an independent sample $x_n(t) \sim f_n(t)$ with a given distribution $f_n(t)$, e.g., Gaussian or binary. Instead, when the source $n$ is inactive, i.e., $s_n(t)=0$, it does not transmit and we have $x_n(t)=0$.

A final remark is in order concerning synchronization requirements. The described model \eqref{eq:signal model} applies even in the absence of time synchronization among the sources, since the vector $\mathbf{x}(t)$ can model a generic sample at discrete time $t$ of the transmitted signals. However, if time synchronization is assumed, and the sources transmit digitally modulated signals, the vector $\mathbf{x}(t)$ can be assumed to contain the constellation points transmitted at the $t$-th symbol period. We note that time and frequency synchronization in sensor networks is a topic of great interest, and it has been investigated in blind scenarios with no pilot symbols as well \cite{Bsync1,Bsync2}. However, the literature about BSS often assumes that sensors and sources are synchronized in order to focus on the technical challenges of BSS \cite{DLWC17,DLWC13,DLWC15,DLWC16}.

\section{Preliminaries: Dictionary Learning} \label{sec:pre}

The DL method proposed in this work leverages prior information about the memory of the sources. We start by reviewing DL methods that do not exploit such information. These methods use only the fact that the signal $\mathbf{x}(t)$ is sparse at any time $t$. Prior information about the memory of each source $x_n(t)$, to be considered in the next section, includes smoothness properties or statistical models.

Assuming only information about the sparseness of $\mathbf{x}(t)$ at each time $t$, a standard approach is to utilize the channel matrix $\mathbf{H}$ as a \textit{dictionary} to be learned to recover $\mathbf{X}$. DL techniques approximate the solution of the maximum likelihood (ML) problem if the fusion center acts as a master clock synchronizing all other nodes.
\begin{equation} \label{eq:DL}
	\underset{\mathbf{H},\mathbf{X} \in \mathcal{X}}{\text{minimize }} ||\mathbf{Y}-\mathbf{H}\mathbf{X}||^2,
\end{equation}
where $\mathcal{X}$ is the set of matrices with sparse columns, that is, with columns containing a limited number of non-zero entries. This problem is not convex with respect to the pair ($\mathbf{H}$, $\mathbf{X}$). DL methods use an iterative procedure, whereby the signal $\mathbf{X} \in \mathcal{X}$ and the channel $\mathbf{H}$ are optimized alternately \cite{DL15}.
In the following, we first discuss solutions for the optimization over the signal $\mathbf{X}$ for a given channel matrix $\mathbf{H}$, and then over the channel $\mathbf{H}$ for a given signal matrix $\mathbf{X}$.

DL methods are subject to inherent permutation and sign ambiguities \cite{Ambi10}. The scaling ambiguity can be instead resolved if one imposes that the channel matrix columns are normalized \cite{MOD99,MDU13,SGK13,Sadeghi13}.

\subsection{Signal Estimation} \label{sec:SE}

For any fixed iterate $\mathbf{H}^{(k)}$ at the $k$-th iteration, from \eqref{eq:DL}, the problem of estimating the signal $\mathbf{X}$ reduces to
\begin{equation} \label{eq:SE}
	\mathbf{X}^{(k+1)}=\underset{\mathbf{X} \in \mathcal{X}}{\text{argmin}} ||\mathbf{Y}-\mathbf{H}^{(k)}\mathbf{X}||^2.
\end{equation}
Standard sparse optimization estimators, such as orthogonal matching pursuit (OMP) \cite[\S 3.1.2]{OMPbook} can be used to address problem \eqref{eq:SE}. Alternatively, one can use the LASSO algorithm \cite{lasso96} to solve the convex problem
\begin{equation} \label{eq:LASSO}
	\underset{{\mathbf{x}(t)}}{\text{minimize }} || \mathbf{y}(t)-\mathbf{H}^{(k)}\mathbf{x}(t)||_2 + \lambda ||\mathbf{x}(t)||_1, \qquad t=1,\dots,T,
\end{equation}
separately for each $t$, where the weight $\lambda$ is a parameter to be determined as a function of the sparsity of vector $\mathbf{x}(t)$.

\subsection{Channel Estimation}

At the $k$-th iteration, for a fixed iterate $\mathbf{X}^{(k+1)}$, the channel estimation step can obtain the next channel iterate $\mathbf{H}^{(k+1)}$ by using different algorithms, such as the Method of Optimal Directions (MOD) \cite{MOD99}, the Multiple Dictionary Update (MDU) \cite{MDU13}, or their enhanced versions proposed in \cite{Sadeghi13}. Here we summarize the enhanced MDU approach, which was shown in \cite{Sadeghi13} to provide the best performance via simulation results.

The MDU approach estimates the channel matrix $\mathbf{H}$ for a given $\mathbf{X}^{(k+1)}$ by following an iterative approach. To elaborate, denote $\mathcal{S}(\mathbf{x})$ the set of indices of the non-zero elements in vector $\mathbf{x}$.
Also, index $(j,k)$ the $j$-th iteration of the MDU algorithm within the $k$-th step of the DL alternate optimization scheme. At the $(j,k)$ iteration, for $j=1,2,\dots$, MDU computes
\begin{equation} \label{eq:MOD}
	\mathbf{H}^{(j,k)}=\mathbf{Y}\mathbf{X}^{(j,k)T}(\mathbf{X}^{(j,k)}\mathbf{X}^{(j,k)T})^{-1},
\end{equation}
and
\begin{equation}  \label{eq:MDU}
	\mathbf{x}^{(j+1,k)}(t)=\mathbf{D}^{(k)}(\mathbf{H}^{(j,k)T}\mathbf{H}^{(j,k)})^{-1}\mathbf{H}^{(j,k)T}\mathbf{y}(t),
\end{equation}
where for all $t$, $\mathbf{D}^{(k)}$ is a diagonal matrix with elements having indices in $\mathcal{S}(\mathbf{x}^{(k+1)}(t))$ equal to $1$ and zero otherwise. The iteration is initialized with $\mathbf{X}^{(1,k)}=\mathbf{X}^{(k+1)}$. Therefore, for a fixed sparsity pattern $\mathcal{S}(\mathbf{x}^{(k+1)}(t))$, MDU alternately estimates channel and signals.

The enhancement proposed in \cite{Sadeghi13} substitutes at iteration $k$ the received signal $\mathbf{Y}$ in \eqref{eq:MOD} and \eqref{eq:MDU} with $\mathbf{Y}^{(k)}=\mathbf{Y}+\mathbf{H}^{(k)}\mathbf{X}^{(k)}-\mathbf{H}^{(k)}\mathbf{X}^{(k+1)}$.

A final remark is that the reviewed DL techniques are designated for flat-fading channels. 
Multipath scenarios would require convolutional dictionary learning \cite{CDL18}, and would be the topic of a future study.

\section{Smooth Dictionary Learning} \label{sec:SDL}

In this section, we propose an improved DL-based source separation algorithm that exploits knowledge about the memory of the sources $n=1,\dots,N$. In particular, we introduce a modified DL scheme that accounts for the intermittent and smooth deterministic model discussed in Sec.~\ref{sec:SM}. 


\begin{figure}
	\centering
	\begin{minipage}[c]{0.7\columnwidth}
		\includegraphics[width=\columnwidth]{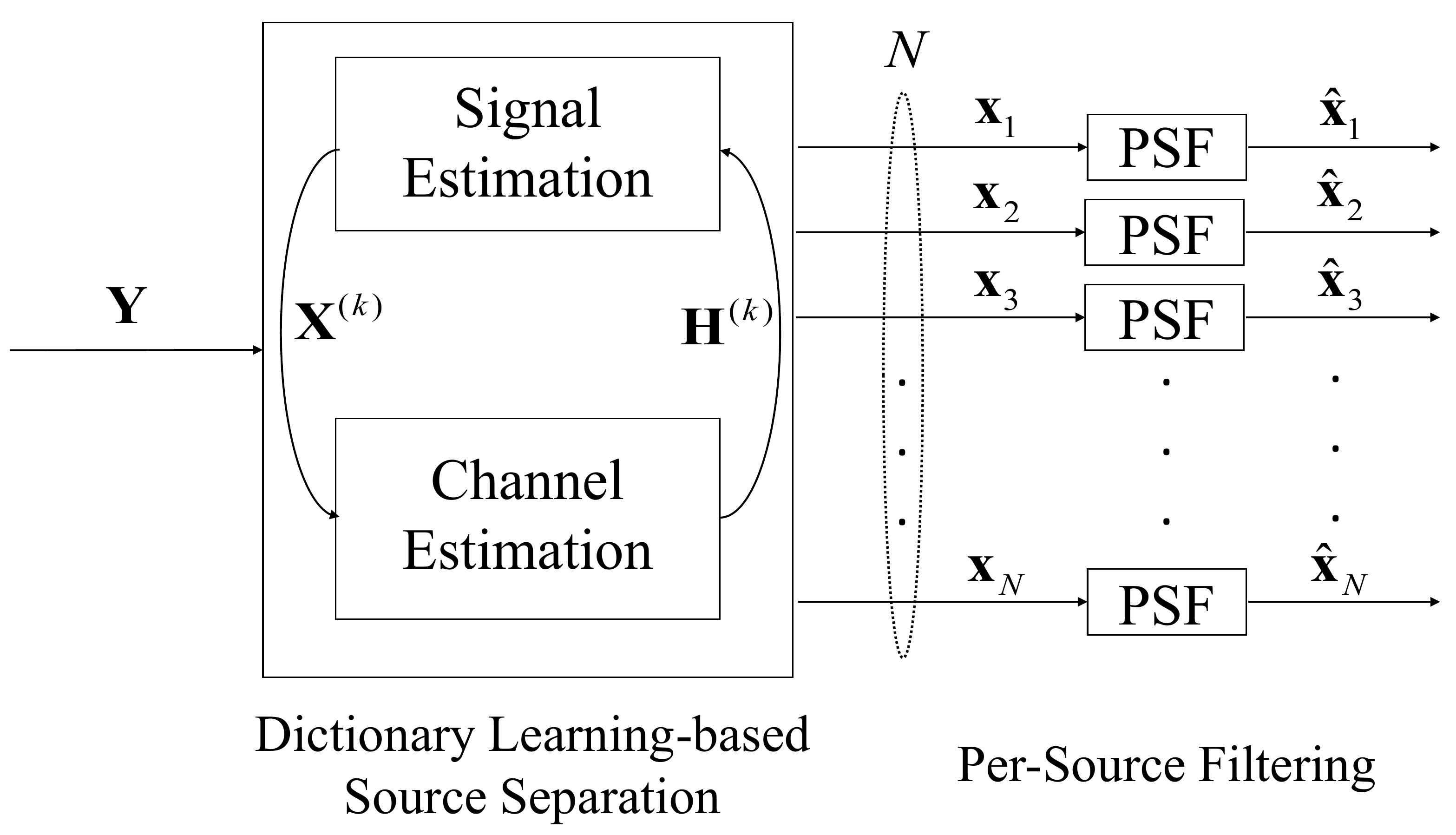}
		\caption{Block diagram of the proposed algorithm with Per-Source Filtering under the hidden Markov model for the sources' activity.}
		\label{fig:AF}
	\end{minipage}
\end{figure}


To account for the assumption of smoothness, the signal estimation step described in Sec.~\ref{sec:SE} is modified by substituting the LASSO algorithm \eqref{eq:LASSO} with a smooth LASSO algorithm \cite{SL11}. Smooth LASSO adds a penalty term for on/off switches within the transmitted signals. This penalty reflects prior knowledge that the transmitted signals do not switch on/off an excessive number of times. Accordingly, at each iteration $k$, given the channel iterate $\mathbf{H}^{(k)}$, instead of using \eqref{eq:LASSO}, the proposed method obtains the updated estimate $\mathbf{X}^{(k)}$ of the channel matrix by solving the problem
\begin{equation} \label{eq:SL}
	\underset{\mathbf{X}}{\text{minimize }}|| \mathbf{Y}-\mathbf{H}^{(k)}\mathbf{X}||^2_2 + \lambda \sum_{t=1}^T ||\mathbf{x}(t)||_1 +\mu \sum_{t=2}^T||\mathbf{x}(t)-\mathbf{x}(t-1)||^2,
\end{equation}
where $\mu$ is a weight parameter that is set depending the level of smoothness expected in the transmitted signals. Larger values of $\mu$ indicate fewer expected changes in the transmitted signal.
Solving \eqref{eq:SL} directly has a complexity in the order of $O(MN^2T^2)$ \cite{CC07}. To reduce the computational complexity, we introduce two approximations of \eqref{eq:SL}, 
namely Sequential Smooth LASSO (SL-SEQ) and Alternating Direction Method of Multipliers-based Smooth LASSO (SL-ADMM). SL-SEQ solves \eqref{eq:SL} column-wise sequentially, and can be found in \cite{SLSEQ05} as an approximation for \eqref{eq:SL}; while SL-ADMM is an ADMM-based \cite{ADMM11,ADMM14} iterative approach we propose to tackle \eqref{eq:SL} with lower computational complexity.

\subsection{Sequential Smooth LASSO (SL-SEQ)}

SL-SEQ solves the problem
\begin{equation} \label{eq:SSL}
	\underset{\mathbf{x}(t)}{\text{minimize }}|| \mathbf{y}(t)-\mathbf{H}^{(k)}\mathbf{x}(t)||^2_2 + \lambda ||\mathbf{x}(t)||_1+\mu ||\mathbf{x}(t)-\mathbf{x}(t-1)||^2,
\end{equation}
sequentially for $t=1,\dots,T$, where $\mathbf{x}(t-1)$ is the solution obtained at the previous step \cite{SLSEQ05}. With this scheme, the quadratic penalty term imposes on the current solution a proximity constraint with respect to the previous one. SL-SEQ has a complexity that in the order of $O(MN^2T)$ as opposed to $O(MN^2T^2)$ since it solves the $T$ problems in \eqref{eq:SSL} sequentially.

\subsection{Alternating Direction Method of Multipliers-Based Smooth LASSO (SL-ADMM)}

We now propose an approximation to Smooth LASSO with ADMM approach \cite{ADMM11,ADMM14}. The novelty of SL-ADMM is to solve BSS problems with lower computational complexity than smooth LASSO for sources that have memory. To this end, we introduce a copy $\mathbf{x}'(t)$ of $\mathbf{x}(t)$ and rewrite problem \eqref{eq:SL} as
\begin{equation} \label{eq:ADMM1}
	\begin{aligned}
		\underset{\mathbf{X},\mathbf{X}'}{\text{minimize}} \text{ } &|| \mathbf{Y}-\mathbf{H}^{(k)}\mathbf{X}||_2 + \lambda \sum_{t=1}^T||\mathbf{x}(t)||_1 + \mu \sum_{t=2}^T||\mathbf{x}(t)-\mathbf{x}'(t-1)||^2, \\
		\text{subject to }& \mathbf{x}'(t)=\mathbf{x}(t), \qquad t=1,\dots,T-1,
	\end{aligned}
\end{equation}
where we have defined the matrix $\mathbf{X}'=[\mathbf{x}'(1),\dots,\mathbf{x}'(T-1)].$
Define the functions $l_1(\mathbf{x}(1))=|| \mathbf{y}(1)-\mathbf{H}^{(k)}\mathbf{x}(1)||_2 + \lambda ||\mathbf{x}(1)||_1$ and $l_t(\mathbf{x}(t),\mathbf{x}'(t-1))=|| \mathbf{y}(t)-\mathbf{H}^{(k)}\mathbf{x}(t)||_2 + \lambda ||\mathbf{x}(t)||_1 + \mu ||\mathbf{x}(t)-\mathbf{x}'(t-1)||^2$, for $t \ge 2$. Then, the augmented Lagrangian \cite{AL69} for problem \eqref{eq:ADMM1} can be written as:
\begin{equation} \label{eq:ADMM2}
	\begin{aligned}
		&\mathcal{L}(\mathbf{x},\mathbf{x}',\boldsymbol{\alpha}) = l_1(\mathbf{x}(1)) + \sum_{t=2}^T l_t(\mathbf{x}(t),\mathbf{x}'(t-1)) + \sum_{t=1}^{T-1} \boldsymbol{\alpha}^T(t) (\mathbf{x}(t)-\mathbf{x}'(t)) + \rho \sum_{t=2}^T||\mathbf{x}(t) - \mathbf{x}'(t-1)||^2 \\
		&= l_1(\mathbf{x}(1))+\boldsymbol{\alpha}^T(1) \mathbf{x}(1) + \sum_{t=2}^{T-1} (l_t(\mathbf{x}(t),\mathbf{x}'(t-1))+\boldsymbol{\alpha}^T(t) \mathbf{x}(t) - \boldsymbol{\alpha}^T(t-1) \mathbf{x}'(t-1)) \\
		&+ l(\mathbf{x}(T),\mathbf{x}'(T-1)) - \boldsymbol{\alpha}^T(T-1) \mathbf{x}'(T-1) + \rho \sum_{t=2}^T||\mathbf{x}(t) - \mathbf{x}'(t-1)||^2,
	\end{aligned}
\end{equation}
where $\boldsymbol{\alpha}(t)$ is the $N \times 1$ Lagrange multipliers for the constraints in \eqref{eq:ADMM1} and $\rho \ge 0$ is a parameter.
The proposed method tackles problem \eqref{eq:ADMM1} via a primal-dual subgradient method that carries out the following steps at each iteration $i$ of ADMM
\begin{itemize}
	\item For the current iterates $\boldsymbol{\alpha}^{(i)}(t)$, solve in parallel the $T$ problems:
	\begin{subequations} \label{eq:ADMMpara}
		\begin{itemize}
			\item[$\cdot$] $\underset{\mathbf{x}(1)}{\text{minimize}} \text{ } l_1(\mathbf{x}(1))+\boldsymbol{\alpha}^{(i)T}(1)\mathbf{x}(1)$, \hfill\refstepcounter{equation}\textup{(\theequation)}
			\item[$\cdot$] $\underset{\mathbf{x}(t),\mathbf{x}'(t-1)}{\text{minimize}} \text{ } l_t(\mathbf{x}(t),\mathbf{x}'(t-1))+\boldsymbol{\alpha}^{(i)T}(t)\mathbf{x}(t)-\boldsymbol{\alpha}^{(i)T}(t-1)\mathbf{x}'(t-1)  +\rho ||\mathbf{x}(t) - \mathbf{x}'(t-1)||^2, \\ \text{ for } t=2,\dots,T-1$, \hfill\refstepcounter{equation}\textup{(\theequation)}
			\item[$\cdot$] $\underset{\mathbf{x}(T),\mathbf{x}'(T-1)}{\text{minimize}} \text{ } l_T(\mathbf{x}(T),\mathbf{x}'(T-1))-\boldsymbol{\alpha}^{(i)T}(T-1)\mathbf{x}'(T-1)+\rho ||\mathbf{x}(T) - \mathbf{x}'(T-1)||^2$, \hfill\refstepcounter{equation}\textup{(\theequation)}
			\item[] obtaining the new iterates $\mathbf{x}^{(i)}(t)$ for $t=1,\dots,T$ and $\mathbf{x}'^{(i)}(t)$ for $t=2,\dots,T$;
		\end{itemize}
	\end{subequations}
	\item Update the Lagrange multipliers as
	\begin{equation}
		\boldsymbol{\alpha}^{(i+1)}(t) \leftarrow \boldsymbol{\alpha}^{(i)}(t)+\rho (\mathbf{x}^{(i)}(t) - \mathbf{x}'^{(i)}(t-1)),
	\end{equation}
\end{itemize}
where $(\cdot)^T$ represents the transpose of a vector or a matrix. 

We emphasize that SL-SEQ has a computational complexity in the order of $O(MN^2T)$, and SL-ADMM has a complexity in the order of $O(MN^2TI)$, where $I$ is the number of iterations in the ADMM approach. In general, $I$ is chosen to be a small value compared to $T$. It is seen that the computational complexity is greatly reduced with SL-SEQ and SL-ADMM compared to solving \eqref{eq:SL} directly.

\section{Per-Source Stochastic Filtering} \label{sec:PSF}

In this section, we propose a further improvement to the DL-based source separation schemes discussed thus far which as shown in Fig.~\ref{fig:AF}, consists of a per-source post-processing step. Specifically, once we have obtained an estimate $\tilde{x}_n(t),t=1,\dots,T$, of the signal of any source $n$ based on one of the DL-based methods discussed above, a perform per-source stochastic filtering (PSF) is applied that leverages the HMM of the sources to remove some of the "noise" in the estimates $\tilde{x}_n(t)$. The "noise" flips the state of the signal from active to inactive and vice-versa. Filtering of the noise leads to a more accurate activity map of the signals.

PSF is based on the hidden Markov model of Fig.~\ref{fig:HMM}. We first consider the case in which the transition probabilities $(p_n, q_n)$ of the hidden Markov model are known, and then study the scenario in which this information is not available.

\subsection{Known Model Parameters}

Adopting a Bayesian formulation, a first approach would be to interpret the signal $\tilde{\mathbf{x}}_n(t)$ obtained from each source $n$ via DL-based source separation as the output of a memoryless channel $p(\tilde{\mathbf{x}}_n|\mathbf{x}_n)$ whose input is the current transmitted signal $\mathbf{x}_n$. Composing this channel with the hidden Markov model of Fig.~\ref{fig:HMM}, one could apply the forward-backward message passing algorithm on the resulting Bayesian network in order to compute the posterior probabilities $p(x_n(t)|\tilde{x}_n(1),\dots,\tilde{x}_n(T))$ for all $t=1,\dots,T$ \cite{BN}. Here we do not follow this approach since it is generally unclear how to define the channel $p(\tilde{\mathbf{x}}_n|\mathbf{x}_n)$, and the resulting message passing algorithm typically requires computationally expensive integrations. In contrast, we take a more pragmatic low-complexity approach, which is described next.

First, we perform a binary quantization of the available estimates $\tilde{x}_n(t),t=1,\dots,T$, by thresholding the absolute value of $\tilde{x}_n(t)$. This provides an estimate $\tilde{s}_n(t)$ of the state $s_n(t)$, since small absolute values of $\tilde{\mathbf{x}}_n(t)$ suggest the absence of signal for source $n$ at time $t$. As a result, we obtain the binary sequence $\tilde{s}_n(t),t=1,\dots,T$, where $\tilde{s}_n(t)=1$ if $|\tilde{x}_n(t)|>\gamma$ for a threshold $\gamma$, and $\tilde{s}_n(t)=0$ otherwise. The selection of the threshold $\gamma$ will allow us to obtain different points on the trade-off between the probability of false alarm and missed detection, as discussed in Sec.~\ref{sec:NR}.

\begin{figure}
	\centering
	\begin{minipage}[c]{0.7\columnwidth}
		\centering
		\includegraphics[width=0.6\columnwidth]{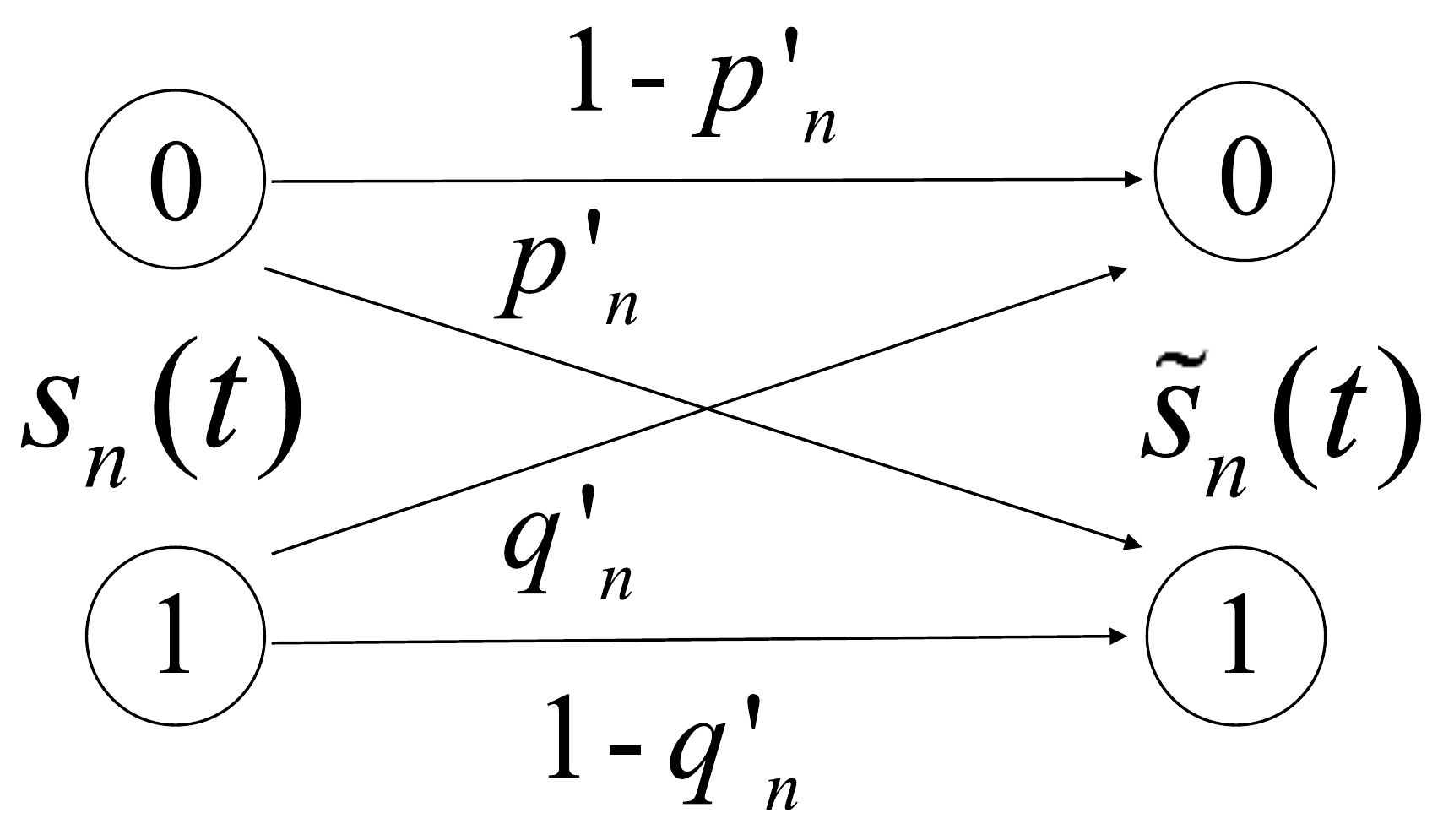}
		\caption{Binary Asymmetric Channel (BAC) describing the relationship between the true state $s_n(t)$ and the observed state $\tilde{s}_n(t)$ obtained from the DL-based source separation step that is assumed by PSF.}
		\label{fig:BAC}
	\end{minipage}
\end{figure}

To formulate the filtering problem, we model the observations $\tilde{s}_n(t),t=1,\dots,T$, as being received at the output of a binary asymmetric channel (BAC) with flip probabilities $p_n'$ and $q_n'$, as illustrated in Fig.~\ref{fig:BAC}. The BAC models the errors made by the DL-based source separation algorithm that produced $\tilde{x}_n(t),t=1,\dots,T$, in detecting the state variables $s_n(t)$. Parameters $p_n'$ and $q_n'$ can be used as additional degrees of freedom in exploring the trade-off points between the probability of false alarm and missed detection. As an alternative, we will discuss in the next subsection how to estimate them from the data.

Given the observations $\tilde{s}_n(t),t=1,\dots,T$, and the parameters $p_n$, $q_n$, $p_n'$ and $q_n'$, we compute the posterior distribution $p(s_n(t)|\tilde{s}_n(1),\dots,\tilde{s}_n(T))$ by using the forward-backward algorithm \cite{ML}. Accordingly, the posterior probability can be written as
\begin{equation} \label{eq:PP}
	\text{Pr}(s_n(t)=1|\tilde{s}_n(1),\dots,\tilde{s}_n(T))=\alpha_t \beta_t,
\end{equation}
where the probability $\alpha_t = \text{Pr}(s_n(t)=1|\tilde{s}_n(1),\dots,\tilde{s}_n(t))$ is obtained as a result of the forward pass, while the probability $\beta_t= \text{Pr}(\tilde{s}_n(t+1),\dots,\tilde{s}_n(T)|s_n(t)=1)$ is obtained from the backward pass as explained in \cite[\S 17.4.3]{ML}. We then estimate $s_n(t)$ using the maximum a posteriori (MAP) approach, i.e., we set $\tilde{s}_n(t)=1$ if the inequality $p(s_n(t)|\tilde{s}_n(1),\dots,\tilde{s}_n(T))>0.5$ holds, and $\tilde{s}_n(t)=0$ otherwise. To recover an estimate $\hat{x}_n(t)$ of the transmitted signal $x_n(t),t=1,\dots,T$, we finally null all entries of $\tilde{x}_n(t)$ corresponding to states $s_n(t)$ that are estimated to be zero, i.e., $\hat{s}_n(t)=0$, while leaving unaltered the values of $\tilde{x}_n(t)$ for times $t$ at which $s_n(t)$ is estimated to be $\hat{s}_n(t)=1$. This can be summarized as
\begin{equation}
	\hat{x}_n(t) =
	\begin{cases}
		\tilde{x}_n(t) & \quad \text{if } \hat{s}_n(t)=1,\\
		0  & \quad \text{if } \hat{s}_n(t)=0.\\
	\end{cases}
\end{equation}

\subsection{Unknown Model Parameters} \label{sec:UMP}

We now study the case in which the parameters $p_n$ and $q_n$ of the HMM are unknown, and need to be estimated. We also jointly estimate the parameters $p_n'$ and $q_n'$ in the BAC of Fig.~\ref{fig:BAC} that is used to derive the forward-backward filtering algorithm. To this end, we apply the Expectation-Maximization (EM) algorithm \cite{ML}, as described next.

Given the estimated states $\tilde{s}_n(t)$ of each source $n$ obtained from the source separation and quantization steps, we would like to estimate the state sequence $s_n(t), \text{ } t=1,\dots,T$, the probabilities $p_n$ and $q_n$ in the transition matrix of the Markov chain, as well as the probabilities $p_n'$ and $q_n'$ in the BAC. The EM algorithm can be detailed as follows.

\begin{itemize}
	
	\item \textbf{Initialization:} Initialize $p^{(0)}_n$, $q^{(0)}_n$, $p'^{(0)}_n$ and $q'^{(0)}_n$.
	
	For each iteration $\nu=1,2,\dots$:
	\item \textbf{E Step}: Given the probabilities $p_n=p^{(\nu-1)}_n$, $q_n=q^{(\nu-1)}_n$, $p'_n=p'^{(\nu-1)}_n$ and $q'_n=q'^{(\nu-1)}_n$, apply the forward-backward algorithm to calculate the probabilities $\alpha_t$ and $\beta_t$ and hence the posterior probabilities \eqref{eq:PP}, for $n=1,\dots,N$.
	
	\item \textbf{M Step}: Update the probability parameters by averaging the sufficient statistics with respect to the posterior distributions identified during the E step \cite{Bishop}. This leads to the updates
	\begin{equation} \label{eq:pn}
		p^{(\nu +1)}_n \leftarrow \frac{\sum_{t=2}^T \text{Pr}(s_n(t)=1,s_n(t-1)=0|\tilde{\mathbf{s}}_n)}{\sum_{t=2}^T \text{Pr}(s_n(t-1)=0|\tilde{\mathbf{s}}_n)},
	\end{equation}
	\begin{equation} \label{eq:qn}
		q^{(\nu +1)}_n \leftarrow \frac{\sum_{t=2}^T \text{Pr}(s_n(t)=0,s_n(t-1)=1|\tilde{\mathbf{s}}_n)}{\sum_{t=2}^T \text{Pr}(s_n(t-1)=1|\tilde{\mathbf{s}}_n)},
	\end{equation}
	where $\tilde{\mathbf{s}}_n=[\tilde{s}_n(1),\dots,\tilde{s}_n(T))]$, $\text{Pr}(s_n(t-1)=0|\tilde{\mathbf{s}}_n)=1-\text{Pr}(s_n(t-1)=1|\tilde{\mathbf{s}}_n)$ is obtained from \eqref{eq:PP}, and the posterior joint probabilities in \eqref{eq:pn} and \eqref{eq:qn} are computed as detailed below; and
	\begin{equation} \label{eq:ppn}
		p'^{(\nu +1)}_n \leftarrow \frac{\sum_{t=1}^T \mathbbm{1}({\tilde{s}_n(t)=1}) \text{Pr}(s_n(t)=0|\tilde{\mathbf{s}}_n)}{\sum_{t=1}^T \text{Pr}(s_n(t)=0|\tilde{\mathbf{s}}_n)},
	\end{equation}
	\begin{equation} \label{eq:qpn}
		q'^{(\nu +1)}_n \leftarrow \frac{\sum_{t=1}^T \mathbbm{1}({\tilde{s}_n(t)=0}) \text{Pr}(s_n(t)=1|\tilde{\mathbf{s}}_n)}{\sum_{t=1}^T \text{Pr}(s_n(t)=1|\tilde{\mathbf{s}}_n)},
	\end{equation}
	where $\mathbbm{1}(a)$ is the indicator function, i.e. $\mathbbm{1}(a)=1$ if $a$ is true and 0 otherwise.
	
	\item \textbf{Stopping Criterion:} The iteration stops when the parameters converge, i.e. $|p^{(\nu)}_n-p^{(\nu-1)}_n|<\epsilon$, where $\epsilon$ is a small value.
	
\end{itemize}

The posterior joint probabilities in \eqref{eq:pn} and \eqref{eq:qn} are computed as \cite{EMHMM01}
\begin{equation} \label{eq:M}
	\text{Pr}(s_n(t)=i,s_n(t-1)=j|\tilde{s}_n(1),\dots,\tilde{s}_n(T))=\alpha_{t-1,j+1} \Phi_{j+1,i+1}^{(\nu)} \text{Pr}(\tilde{s}_n(t)|s_n(t)=i) \beta_{t,i+1}, \qquad i,j=0,1,
\end{equation}
where $\alpha_{t,j}$ is the $j$-th element of vector $\boldsymbol{\alpha}_t=[1-\alpha_t,\alpha_t]^T$; $\beta_{t,i}$ is the $i$-th element of vector $\boldsymbol{\beta}_t=[1-\beta_t,\beta_t]^T$; $\Phi_{j,i}^{(\nu)}$ is the $(j,i)$-th element of the transition matrix $\boldsymbol{\Phi}^{(\nu)}_n$ in the $\nu$-th iteration; and matrix $\boldsymbol{\Phi}^{(\nu)}_n$ is defined as
\begin{equation}
	\boldsymbol{\Phi}^{(\nu)}_n=
	\begin{bmatrix}
		1-p^{(\nu)}_n & p^{(\nu)}_n \\
		q^{(\nu)}_n & 1-q^{(\nu)}_n
	\end{bmatrix}.
\end{equation}

\section{Numerical Results} \label{sec:NR}
	
In this section, we present numerical results to obtain insights into the performance of different DL-based source separation schemes and on the advantage of PSF. We consider the following source separation schemes, all implemented with and without the post-processing PSF step: OMP, LASSO, SL-SEQ and SL-ADMM. 
As performance criteria, we adopt the probability of false alarm $P_{fa}$ and probability of detection $P_d$. The detection probability $P_d$ is the ratio between the number of correctly detected active sources and the total number of active sources over the $T$ symbols, and the false alarm probability $P_{fa}$ is the ratio between the number of incorrectly detected active sources and the total number of inactive sources over the $T$ time samples. We also consider the performance of source estimation in terms of error vector magnitude (EVM) to synchronized digital transmission (see Sec.~\ref{sec:SM}). 

Unless stated otherwise, the numerical results were obtained for $N=30$ sources, $M=20$ sensors, signal to noise ratio (SNR) per source per sample 30 dB, $T=1000$ time samples. We assume an HMM for the state $s_n(t)$, which is defined by transition probabilities $p_n=0.0022$, $q_n=0.02$, for all $N$ sources, so that an average $Np_n/(p_n+q_n)=3$ sources are active at each time sample $t$, and the average duration of transmission is $1/q_n=50$ time samples. Also, unless stated otherwise, the algorithms with PSF are implemented with fixed values $p'_n=0.02$ and $q'_n=0.27$. We optimize numerically over the multipliers $\lambda$ and $\mu$ in order to satisfy given constants on the probability of detection $P_d$ or the probability of false alarm $P_{fa}$. For PSF with unknown model parameters, we initialize the transition probabilities of the Markov model as $p_n=0.5$, $q_n=0.5$, and the flip probabilities of the BAC as $p'_n=0.1$, $q'_n=0.2$. Throughout this section, the number of iterations for SL-ADMM was fixed at $K=30$; the Lagrange multipliers $\boldsymbol{\alpha}$ in \eqref{eq:ADMMpara} are initialized as the all-one vector; $\rho$ is set as 0.1 and $\gamma$ is fixed at 0.5.

\begin{figure}
	\centering
	\begin{minipage}[c]{0.7\columnwidth}
		\centering
		\includegraphics[width=\columnwidth]{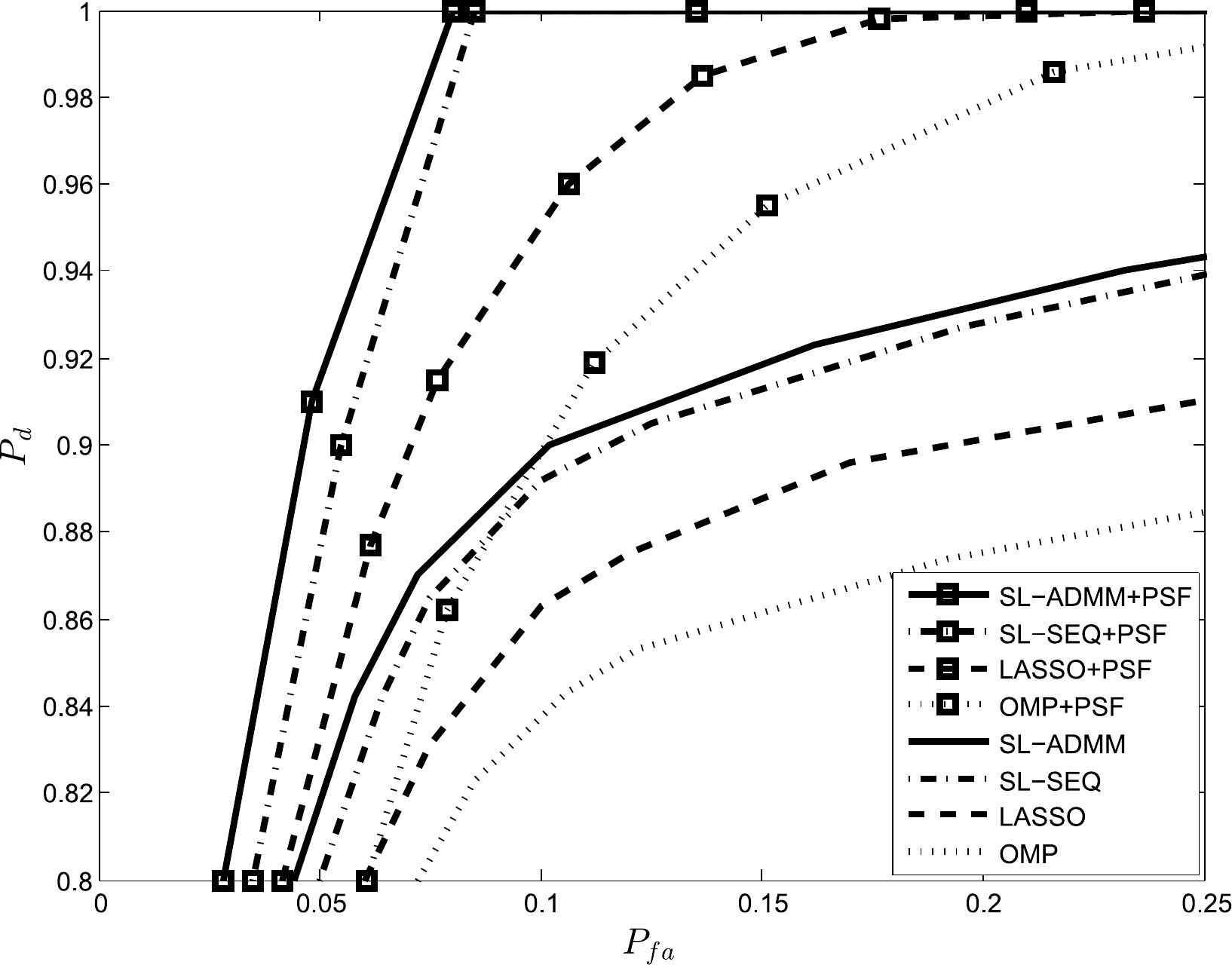}
		\caption{Probability of detection $P_d$ versus probability of false alarm $P_{fa}$ for the considered algorithms ($N=30$, $M=20$, $T=1000$, SNR$=30$ dB, $p_n=0.0022$, $q_n=0.02$).}
		\label{fig:Comp}
	\end{minipage}
\end{figure}

\subsection{Tuning Multipliers $\lambda$ and $\mu$}

We start by discussing the choice of the multipliers $\lambda$ and $\mu$ in optimization problems \eqref{eq:SSL} and \eqref{eq:ADMM1} for SL-SEQ and SL-ADMM. These parameters penalize the number of non-zero elements in the solution vector and the number of state changes of each source, respectively. 
The choice of these two multipliers is scenario-dependent, and a discussion on their optimal selection can be found in \cite{GENR13}. In our extensive simulations, we found that the rule-of-thumb choices $\lambda = 1/$SNR and $\mu =0.1/q_n$, where $q_n$ is the transition probability from an active to an inactive state (see Fig.~\ref{fig:HMM}), works well in practice. The selection $\lambda = 1/$SNR reflects a decrease in the relevance of the sparsity prior as the quality of the observations increases \cite{GENR13}. In contrast, the multiplier $\mu \propto 1/q_n$ increases as the average transmission duration $1/q_n$ increases, implying fewer state changes.

\begin{figure}
	\centering
	\begin{minipage}[c]{0.7\columnwidth}
		\centering
		\includegraphics[width=\columnwidth]{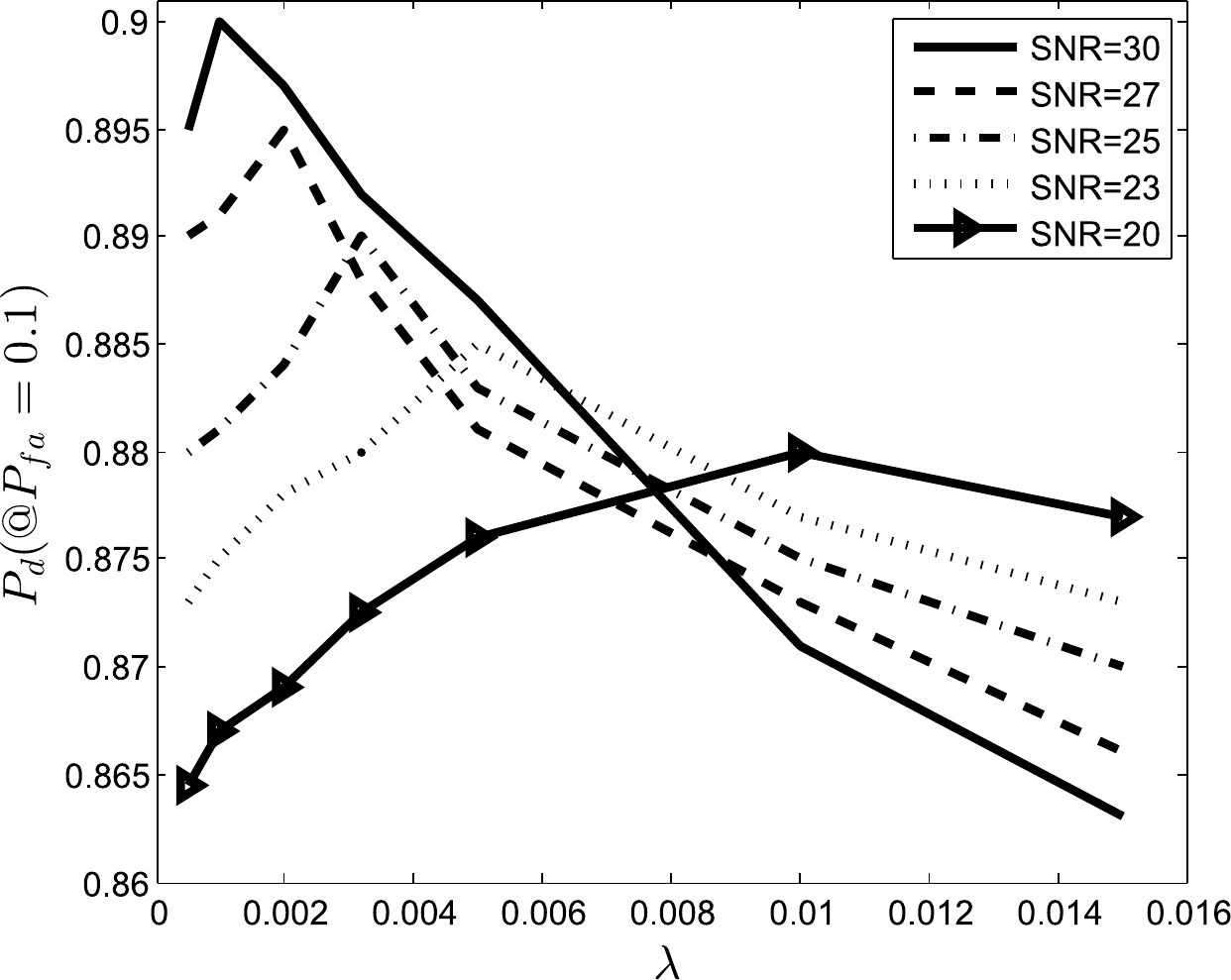}
		\caption{Probability of detection $P_d$ when the probability of false alarm is $P_{fa}=0.1$ versus $\lambda$ for the SL-ADMM algorithm ($N=30$, $M=20$, $T=1000$, $p_n=0.0022$, $q_n=0.02$).}
		\label{fig:lambda}
	\end{minipage}
\end{figure}

\begin{figure}
	\centering
	\begin{minipage}[c]{0.7\columnwidth}
		\centering
		\includegraphics[width=\columnwidth]{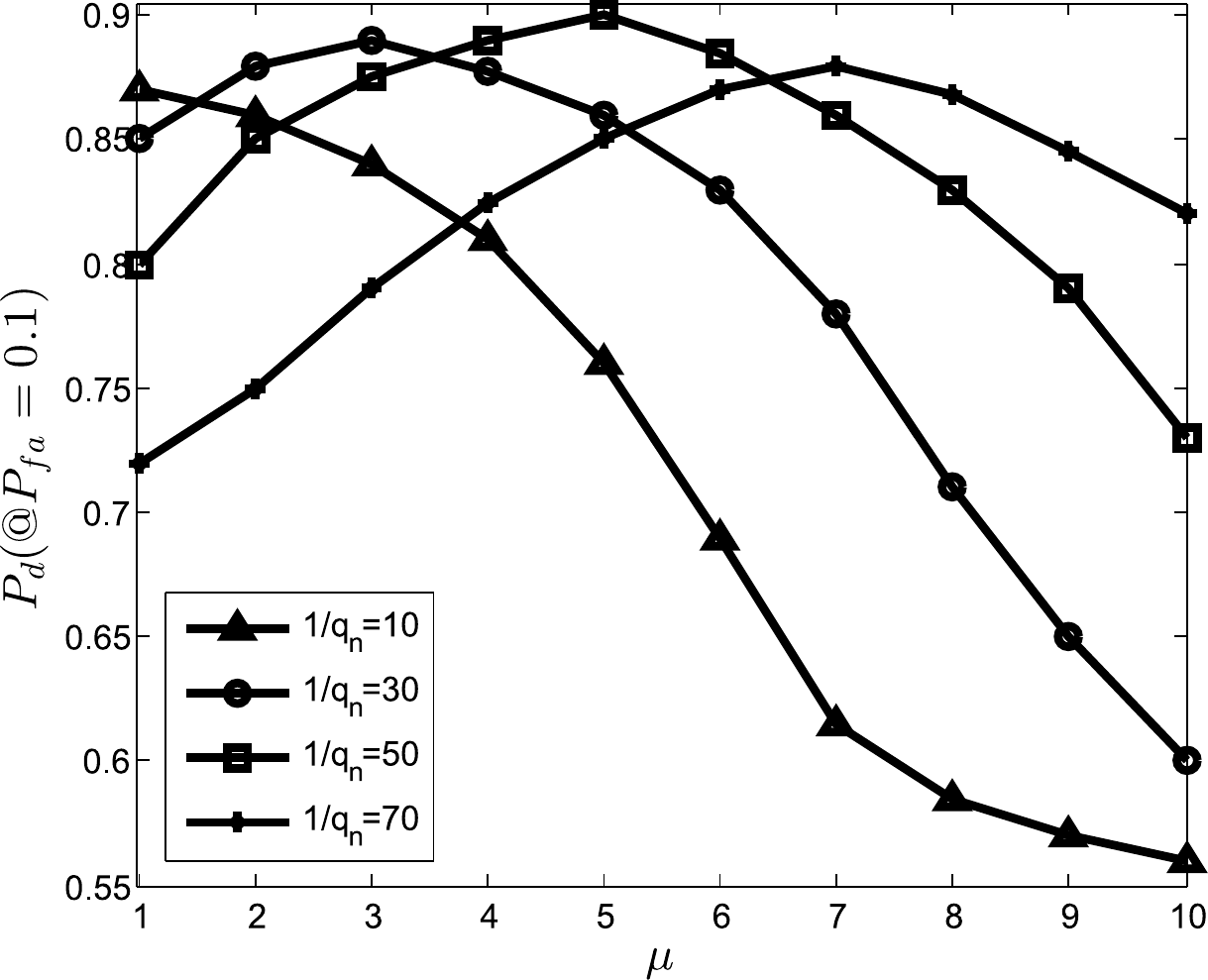}
		\caption{Probability of detection $P_d$ when the probability of false alarm is $P_{fa}=0.1$ versus $\mu$ for the SL-ADMM algorithm ($N=30$, $M=20$, $T=1000$, SNR$=30$ dB.}
		\label{fig:mu}
	\end{minipage}
\end{figure}

As an example, Fig.~\ref{fig:lambda} and Fig.~\ref{fig:mu} show the probability of detection $P_d$ for a fixed probability of false alarm $P_{fa}=0.1$ versus $\lambda$ and $\mu$, respectively. From Fig.~\ref{fig:lambda}, we observe that, when varying the SNR level, the optimal selection of $\lambda$ changes according to the mentioned inverse proportionality rule $1/$SNR. In a similar manner, Fig.~\ref{fig:mu} confirms the validity of our choice for multiplier $\mu$.

\subsection{Source Activity Detection}

We then investigate the trade-off between the probability of detection $P_d$ versus the probability of false alarm $P_{fa}$ in Fig.~\ref{fig:Comp}. The transmitted signal $x_n(t)$ is assumed here to be distributed as $x_n(t) \sim \mathcal{N}(0,1)$ whenever $s_n(t)=1$. We also assume known source statistics $p_n$ and $q_n$. A first observation is that both SL-SEQ and SL-ADMM significantly outperform all other source separation schemes, with the latter obtaining some performance gain over the former. Furthermore, PSF provides a significant performance boost for all algorithms. For example, at $P_{fa}=0.07$, the probability of detection with SL-ADMM is increased from 0.87 to 0.98.

\begin{figure}
	\centering
	\begin{minipage}[c]{0.7\columnwidth}
		\centering
		\includegraphics[width=\columnwidth]{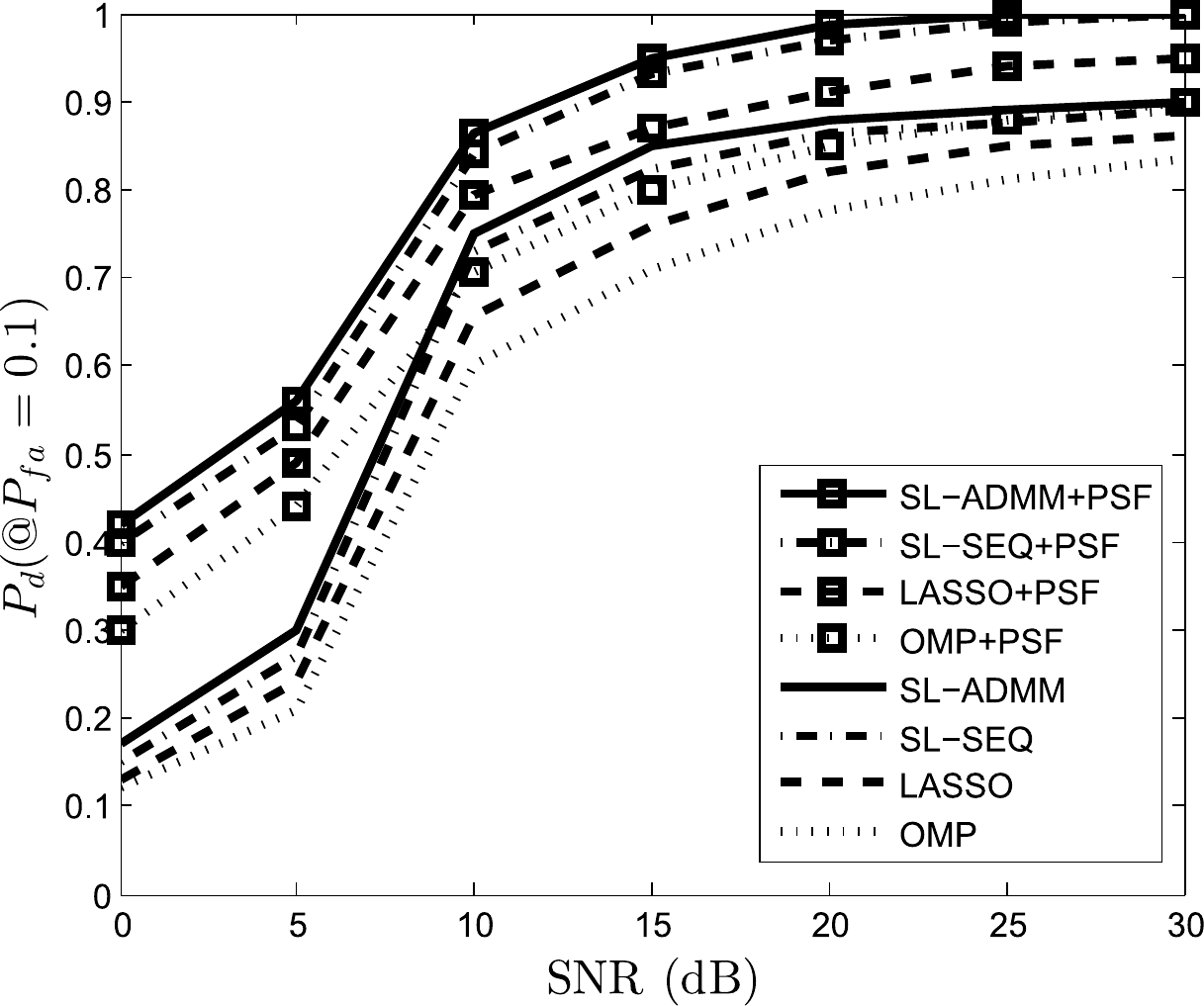}
		\caption{Probability of detection $P_d$ when the probability of false alarm is $P_{fa}=0.1$ versus SNR for the considered algorithms ($N=30$, $M=20$, $T=1000$, $p_n=0.0022$, $q_n=0.02$).}
		\label{fig:SNR}
	\end{minipage}
\end{figure}

\begin{figure}
	\centering
	\begin{minipage}[c]{0.7\columnwidth}
		\centering
		\includegraphics[width=\columnwidth]{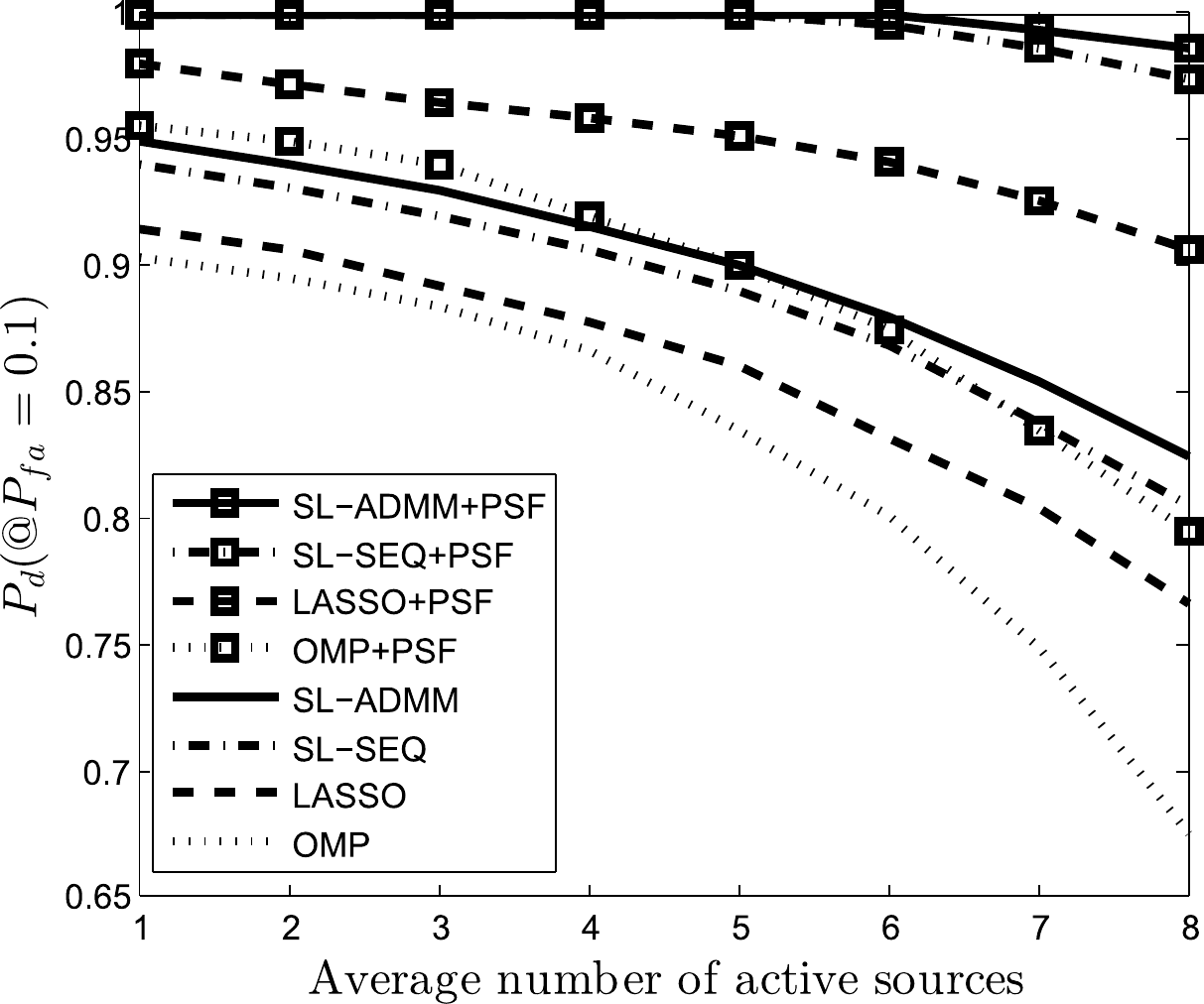}
		\caption{Probability of detection $P_d$ when the probability of false alarm is $P_{fa}=0.1$ versus average number of active sources ($N=30$, $M=20$, $T=1000$, SNR$=30$ dB, $q_n=0.02$).}
		\label{fig:SP}
	\end{minipage}
\end{figure}

We now explore the impact of different operating regimes on the probability of detection when the probability of false alarm is constrained to be smaller than 0.1. In particular, Fig.~\ref{fig:SNR} shows performance with respect to different SNR values, and Fig.~\ref{fig:SP} investigates the impact of the sparsity level, which is defined as the ratio between the average number $Np_n/(p_n+q_n)$ of active sources and the total number $N$ of sources. The former is modified by changing $p_n$. Both figures confirm the main conclusions obtained above in terms of the relative performance of the considered schemes. Moreover, Fig.~\ref{fig:SNR} suggests that the performance undergoes a threshold phenomenon with respect to the SNR, particularly if implemented without PSF. It is also seen that SL with PSF is able to obtain a vanishing probability of missed detection as the SNR increases, unlike the other schemes whose probability of detection reaches a ceiling lower than one. Finally, Fig.~\ref{fig:SP} indicates that SL with PSF is robust to the sparsity level, while the other schemes are extremely sensitive to an increase in the average number of active sources.

\subsection{Signal Estimation}

\begin{figure}
	\centering
	\begin{minipage}[c]{0.7\columnwidth}
		\centering
		\includegraphics[width=\columnwidth]{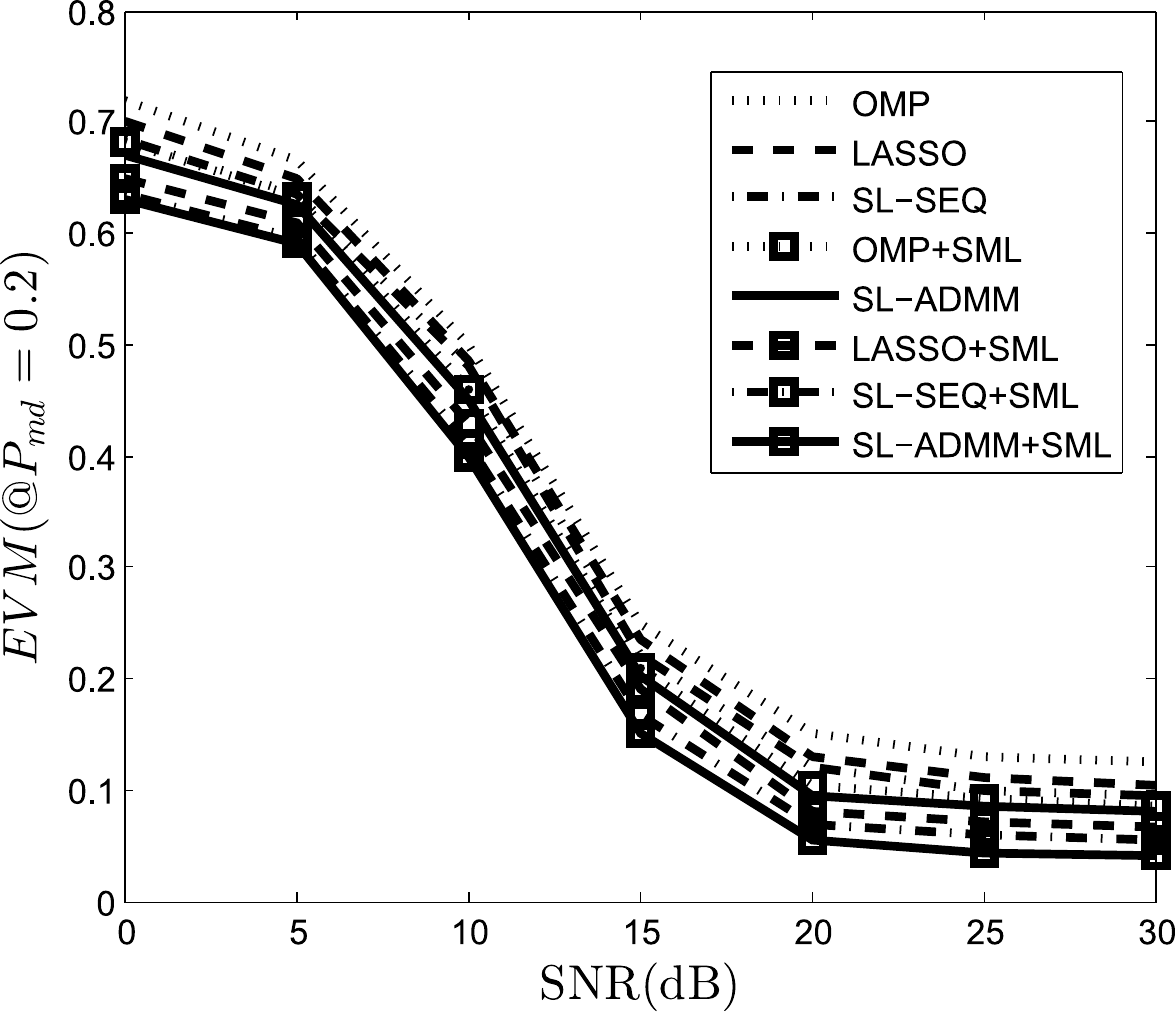}
		\caption{EVM versus SNR ($N=30$, $M=20$, $T=1000$, $p_n=0.0022$, $q_n=0.02$).}
		\label{fig:EVM}
	\end{minipage}
\end{figure}

As seen above, the SL schemes with PSF have the best performance in terms of source activity detection. Here, we study the performance in terms of the quality of signal estimation. To this end, we adopt the criterion of the EVM. EVM is defined as
\begin{equation}
	\text{EVM}(\%)=\sqrt{\frac{\sum_{n=1}^{N} \sum_{t \in \mathcal{D}_n} (x_n(t)-\tilde{x}_n(t))^2}{\sum_{n=1}^{N} \sum_{t \in \mathcal{D}_n} x_n^2(t)}} \times 100\%,
\end{equation}
where $\mathcal{D}_n$ is the set of time samples in which user $n$ is correctly detected as active. Here we assume that, when $s_n(t)=1$, the transmitted signal is binary, that is $x_n(t)=$1 or $-1$ with equal probability.


Assuming again known sources' transition probabilities, Fig.~\ref{fig:EVM} shows the EVM for all the considered algorithms with respect to SNR where the probability of missed detection $P_{md}$ is constrained to be smaller than 0.2. The behavior of the EVM is in line with the discussion above regarding source activity detection. In particular, we observe a 40\% decrease at 30 dB SNR that are achievable with SL and PSF, as well as the threshold behavior as a function of the SNR.

\subsection{Parameter Estimation}

\begin{figure}
	\centering
	\begin{minipage}[c]{0.7\columnwidth}
		\centering
		\includegraphics[width=\columnwidth]{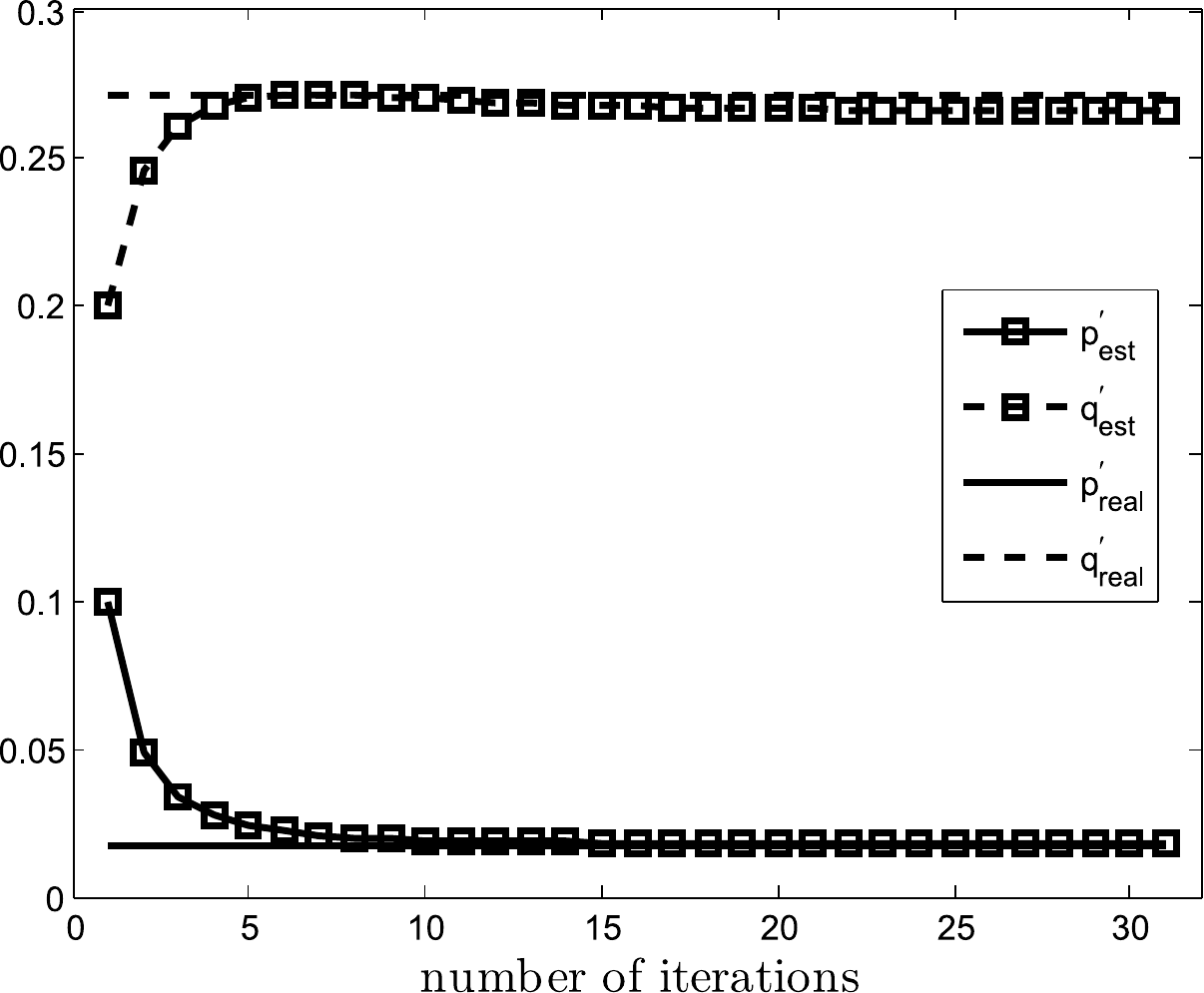}
		\caption{Estimation performance of $p'_n$ and $q'_n$ for BAC in the EM algorithm ($N=30$, $M=20$, $T=1000$, SNR$=30$ dB, $p_n=0.0022$, $q_n=0.02$).}
		\label{fig:HP}
	\end{minipage}
\end{figure}

\begin{figure}
	\centering
	\begin{minipage}[c]{0.7\columnwidth}
		\centering
		\includegraphics[width=\columnwidth]{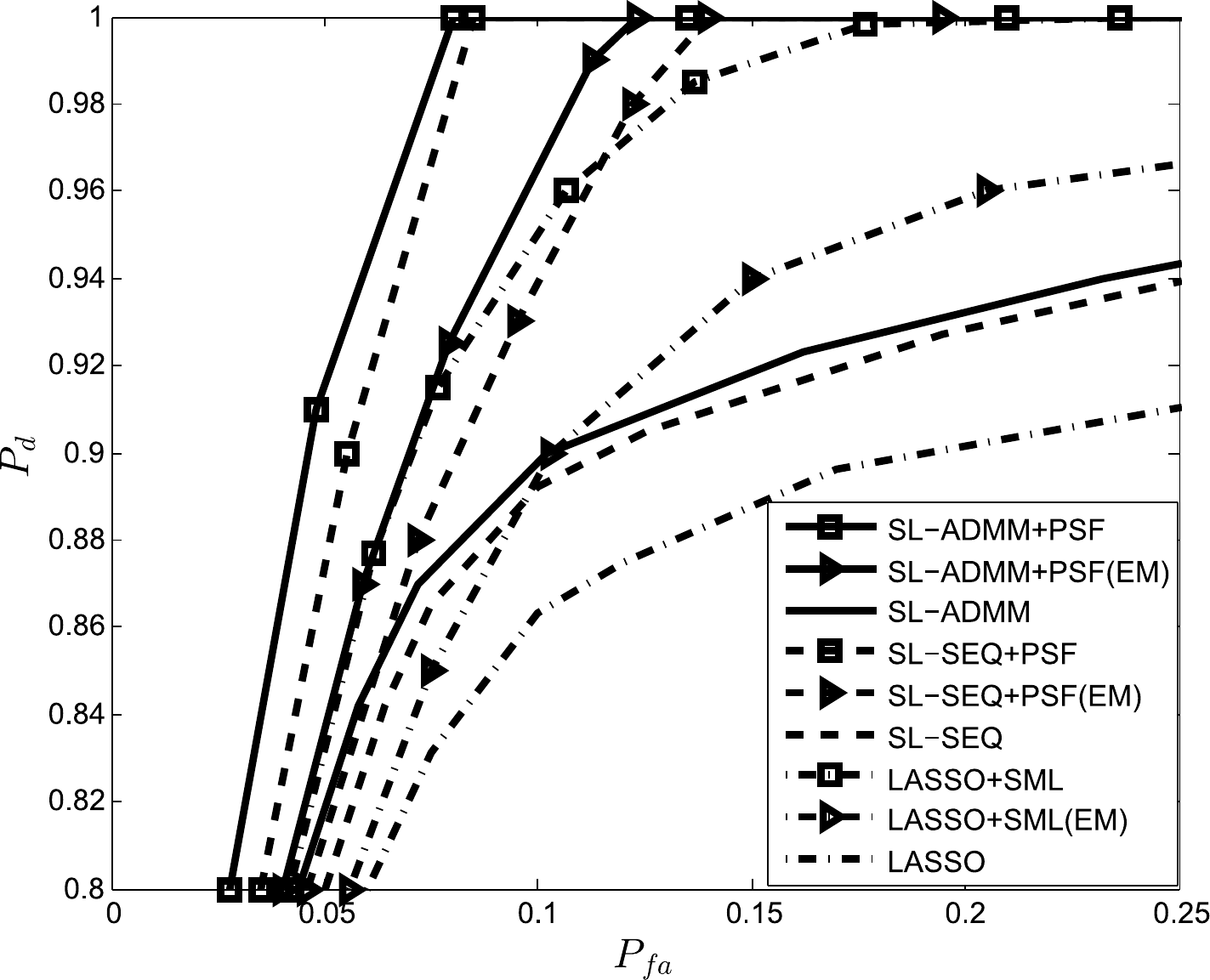}
		\caption{Probability of detection $P_d$ versus probability of false alarm $P_{fa}$ for the considered algorithms with EM algorithm ($N=30$, $M=20$, $T=1000$, SNR$=30$ dB, $p_n=0.0022$, $q_n=0.02$).}
		\label{fig:EM}
	\end{minipage}
\end{figure}

Here, we evaluate the performance loss incurred when the parameters $p_n$ and $q_n$ in the Markov model defining the sources' activities, are not known. Using the framework of Sec.~\ref{sec:UMP}, We jointly estimate the parameters $p_n$, $q_n$, $p'_n$ and $q'_n$ with the EM algorithm. 
Fig.~\ref{fig:HP} investigates the performance of the estimation of the hidden parameters $p'_n$ and $q'_n$ in the BAC. From the figure, the estimate is seen to be close to the real value. The parameters $p_n$ and $q_n$ in the HMM have similar estimation accuracy as $p'_n$ and $q'_n$.
Fig.~\ref{fig:EM} shows the probability of detection $P_d$ versus the probability of false alarm $P_{fa}$ under the same conditions as Fig.~\ref{fig:Comp}. A first observation is that, even with unknown source parameters, PSF can provide a significant performance boost. Nevertheless, the gain is somewhat reduced as compared to the case with perfect knowledge. For instance, for SL-ADMM, when $P_{fa}=0.07$, the probability of detection $P_d$ is reduced from 0.98 to 0.89. The loss is generally increases when $T$ is smaller.

\section{Conclusions}

In this paper, we have introduced a two-stage Dictionary Learning (DL)-based algorithm for solving the BSS problem as the presence of radio sources with memory that are observed over slow flat-fading wireless channels. The DL stage of the proposed algorithm exploits source memory information to aid with the source separation. The effect of source memory is accounted for by a penalty term that discourages short-duration transmissions. The PSF stage utilizes source model information to learn about unknown source model parameters to further enhance source estimation. Numerical results show that the proposed algorithm outperforms existing DL algorithms in terms of source activity detection as well as source signal estimation. As a representative numerical example, even with unknown source statistics, it is shown that with 30 potential sources, 20 sensors and average 3 sources active in each time sample, it is possible to increase the probability of detection from 0.9 without PSF to 0.96 with PSF at 0.1 probability of false alarm and 30 dB SNR. The proposed algorithm can also decrease the EVM from by 40\% at 0.8 probability of detection at 30 dB SNR. In addition, the proposed algorithm is robust to the source sparsity level, while existing DL algorithms are extremely sensitive to an increase in the number of active sources.

\bibliographystyle{IEEEtran}
\bibliography{IEEEabrv,references}

\end{document}